\documentclass[12pt]{iopart}
\usepackage{iopams}
\usepackage{epsfig}
\eqnobysec
\newcommand{\AI}{A{\rm I}}
\newcommand{\AII}{A{\rm I\!I}}
\newcommand{\AIII}{A{\rm I\!I\!I}}
\newcommand{\CI}{C{\rm I}}
\newcommand{\CII}{C{\rm I\!I}}
\newcommand{\BDI}{BD{\rm I}}
\newcommand{\DIII}{D{\rm I\!I\!I}}
\begin{document}
\title[Bott-Kitaev Periodic Table and the Diagonal Map]
{Bott-Kitaev Periodic Table and the Diagonal Map}
\author{R.\ Kennedy and M.R.\ Zirnbauer}
\address{Institut f\"ur Theoretische Physik \\ Universit\"at zu
K\"oln \\ Z\"ulpicher Stra{\ss}e 77 \\ 50937 K\"oln \\ Germany}
\begin{abstract}
Building on the 10-way symmetry classification of disordered fermions, the authors have recently given a homotopy-theoretic proof of Kitaev's ``Periodic Table'' for topological insulators and superconductors. The present paper offers an introduction to the physical setting and the mathematical model used. Basic to the proof is the so-called Diagonal Map, a natural transformation akin to the Bott map of algebraic topology, which increases by one unit both the momentum-space dimension and the symmetry index of translation-invariant ground states of gapped free-fermion systems. This mapping is illustrated here with a few examples of interest. (Based on a talk delivered by the senior author at the Nobel Symposium on ``New Forms of Matter: Topological Insulators and Superconductors''; Stockholm, June 13-15, 2014).
\end{abstract}
\maketitle

\section{Introduction}

The main subject of this paper is Kitaev's ``Periodic Table'' for topological insulators and superconductors \cite{kitaev}. Put forward some six years ago, it organizes into a systematic scheme all of the ``stable'' topological states of symmetry-protected gapped free-fermion systems; it encompasses the integer quantum Hall effect, the quantum spin Hall insulator, the Fu-Kane-Mele $\mathbb{Z}_2$ topological insulator, the Majorana chain, superfluid ${}^3$He in the B-phase, etc., as well as some further states that are yet to be realized in experiment. For the reader's convenience, the ``real'' sub-table of Kitaev's Periodic Table is reproduced in Table \ref{fig:1}, in the particular form we deem most appropriate. Its most striking feature is an 8-fold periodicity. A certain amount of foundational work on it has been done and, in particular, efforts have been made to unveil the principle behind the periodic structure. The most important papers in this regard are \cite{SRFL2008,kitaev,TeoKane,StoneEtAl,FHNQWW,AK,FreedMoore}.

Now the constancy of the entries along the diagonal of the Periodic Table begs the question: is there a so-called ``Diagonal Map'', which takes a symmetry-protected topological phase in $d$ dimensions and transforms it into another such phase in one dimension higher and in the neighboring symmetry class? Note that this map applied to the Majorana chain should give a two-dimensional time-reversal invariant superconductor, and applied again, a three-dimensional $\mathbb{Z}_2$ topological insulator. Starting from the Kitaev chain, its application should give a two-dimensional chiral $p$-wave superconductor, and applied again something like ${}^3$He-B; and so on.

Such a mapping was actually written down by Teo and Kane some time ago. In an appendix to \cite{TeoKane} they give two formulas. The first one specifies how to go from a chiral Hamiltonian to a non-chiral one. The second formula takes you from a non-chiral Hamiltonian back to a chiral one by tensoring with a quasi-spin degree of freedom, and it comes with some case-dependent instructions as to which Pauli matrices to use.

Here we ask the same question but with a higher level of ambition. (i) Is there a ``master'' diagonal map that handles all cases at once, by a single and universal principle? (ii) Can one make a convincing argument that the map indeed gives a one-to-one correspondence between symmetry-protected topological phases? (As mathematical physicists, we would like to formulate and prove a theorem.) (iii) Can one specify the precise conditions under which the Periodic Table applies, and when it does not?

Our modest goal in this short contribution is to convince the reader that such a map does exist, and to explain how it works. (It is described in more detail in \cite{RK-MZ}.)

\begin{table}
\begin{center}
\begin{tabular}{|c|c|c|c|c|c|c|c|}
\hline
$s$	
& $C_s(8r)$
& $R_s(8r)$
& $0d$
& $1d$
& $2d$
& $3d$
& class \\
\hline
0
& $\cup_{p+q=16r}\, \mathrm{U}_{16r}/ (\mathrm{U}_{p} \times \mathrm{U}_{q})$
& $\mathrm{O}_{16r}/ \mathrm{U}_{8r}$
& ${\mathbb Z}_2$		
& ${\mathbb Z}_2$	
& ${\mathbb Z} $	
& $0$
& $D$ \\
1 	
& $(\mathrm{U}_{8r} \times \mathrm{U}_{8r}) / \mathrm{U}_{8r}$
& $\mathrm{U}_{8r}/\mathrm{Sp}_{8r}$
& $0$				
& ${\mathbb Z}_2$
& ${\mathbb Z}_2$	
& ${\mathbb Z}$
& $\DIII$ \\
2  	
& $\cup_{p+q=8r}\, \mathrm{U}_{8r}/(\mathrm{U}_{p}\times \mathrm{U}_{q})$
& $\cup_{p+q=4r} \, \mathrm{Sp}_{8r}/(\mathrm{Sp}_{2p}\times \mathrm{Sp}_{2q})$
& ${\mathbb Z}_{4r+1}$		
& $0$			
& ${\mathbb Z}_2$	
& ${\mathbb Z}_2$
& $\AII$ \\
3 	
& $(\mathrm{U}_{4r} \times \mathrm{U}_{4r}) / \mathrm{U}_{4r}$
& $(\mathrm{Sp}_{4r} \times \mathrm{Sp}_{4r}) / \mathrm{Sp}_{4r}$
& $0$				
& ${\mathbb Z}$ 	
& $0$ 			
& ${\mathbb Z}_2$
& $\CII$ \\
4
& $\cup_{p+q=4r}\, \mathrm{U}_{4r}/(\mathrm{U}_{p}\times \mathrm{U}_{q})$
& $\mathrm{Sp}_{4r}/\mathrm{U}_{2r}$
& $0$				
& $0$ 			
& ${\mathbb Z}$ 	
& $0$
& $C$ \\
5 	
& ($\mathrm{U}_{2r} \times \mathrm{U}_{2r}) / \mathrm{U}_{2r}$
& $\mathrm{U}_{2r}/\mathrm{O}_{2r}$
& $0$				
& $0$ 			
& $0$			
& ${\mathbb Z}$
& $\CI$ \\
6  	
& $\cup_{p+q=2r}\, \mathrm{U}_{2r}/ (\mathrm{U}_{p}\times \mathrm{U}_{q})$
& $\cup_{p+q=2r}\, \mathrm{O}_{2r}/ (\mathrm{O}_{p}\times \mathrm{O}_{q})$
& ${\mathbb Z}_{2r+1}$		
& $0$			
& $0$ 			
& $0$
& $\AI$ \\
7
& ($\mathrm{U}_{r} \times \mathrm{U}_{r}) / \mathrm{U}_r$
& ($\mathrm{O}_{r} \times \mathrm{O}_{r}) / \mathrm{O}_r$
& ${\mathbb Z}_2$		
& ${\mathbb Z}$	
& $0$			
& $0$
& $\BDI$ \\
\hline
\end{tabular}
\caption{Bott-Kitaev Periodic Table for topological insulators and superconductors (more precisely, the ``real'' sub-table thereof). The first column specifies the number $s$ of real pseudo-symmetries, the second and third column spell out the corresponding classifying spaces as defined in Eqs.\ (\ref{eq:3.7}) and (\ref{eq:3.8}). (These are the same symmetric spaces that appear in the Tenfold Way of disordered fermions \cite{HHZ}.) Columns four to seven list the sets of homotopy classes of ground-state vector bundles of class $s$ (as defined in Section \ref{sect:3.2}) as a function of the momentum-space dimension between $d = 0$ and $d = 3$. Note that these are really just sets, not groups. The table assumes that the momentum space is a sphere and that $d \ll r$. (Precise bounds on $d$ versus $r$ are derived in \cite{RK-MZ}, and the case of a momentum torus is treated in \cite{KG14}.) The last column gives the Cartan symmetry class of the disordered free-fermion Hamiltonian.} \label{fig:1}
\end{center}
\end{table}

\section{Distinctive features of our approach}

Here are the highlights and special points that distinguish our approach \cite{RK-MZ} from what is commonly done in the published literature.

First of all, we start from the tenet that a symmetry is a unitary or anti-unitary transformation that \emph{commutes} with the Hamiltonian. To make sure that this point is taken, let us emphasize that the common lore advertising the Periodic Table makes ample use of operations that \emph{anti-commute} with the Hamiltonian. For a number of reasons we do {\bf not} accept such operations as (true) symmetries. For one, they produce very highly excited states when applied to the ground state. For another, if one relaxes the condition that symmetries commute with the Hamiltonian, then why should one allow operations that anti-commute with it but forbid more general relations of, say, parafermionic or quantum-group type? Yet another reason is that the common lore gives no clue as to why the symmetry classes are arranged in the particular sequence they are. In our work, armed with the tight notion of symmetry, we deduce this sequence from first principles, so to speak.

Secondly, to the extent that only the static properties (as opposed to the dynamical response) of the physical system are under investigation, the classification problem at hand is a problem of classifying \emph{ground states}. Therefore, once the symmetry class has been determined, the Hamiltonian leaves the scene and does not reappear in our approach. Thus we work directly with a mathematical model for the ground state and, in particular, we have no need for the commonly invoked process of ``spectral flattening'' of the Hamiltonian.

Thirdly, and most importantly, our work follows a different principle of topological classification than usual. Starting with Kitaev, the community has largely relied on the algebraic tools of $K$-theory to define and compute topological invariants. In contrast, in our work we use tools from homotopy theory. Let it be stressed that homotopy classes are finer and carry more information than do $K$-theory classes, in general.

Last but not least, we work in the standard framework of Hermitian quantum mechanics over the complex numbers. While this may sound like a ``no-brainer'', there actually exist claims in the literature that one is better off working over the real numbers, like Dyson did in his Threefold Way. The thinking behind this was presumably that the 8-fold periodicity of (the real sub-table of) Kitaev's Table is reminiscent of a periodicity phenomenon for real Clifford algebras. While that is certainly true, it turned out to be most revealing for us to keep the real structure flexible. In fact, to get the best perspective of the Diagonal Map, we need to invoke two different operations of taking the complex conjugate.

\section{Universal Model for Free-Fermion Ground States}

The bulk of this paper consists of two parts. The second part, introducing the Diagonal Map, is based on the first part, which describes a universal model for free-fermion ground states of gapped systems with symmetries.

\subsection{Ground states as vector bundles}

We use the standard formulation of second quantization, denoting fermion annihilation operators by $c$ and creation operators by $c^\dagger$. The symbol $M$ stands for momentum space and momenta are denoted by $k$. (For simplicity, we assume translation invariance for now and comment on the disordered situation later.)

It is a basic fact of many-body theory that any translation-invariant free-fermion ground state $| {\rm g.s.} \rangle$ is uniquely determined by specifying for each momentum $k \in M$ the quasi-particle operators $\widetilde{c}_1(k), \ldots, \widetilde{c}_n(k)$ that annihilate it:
\begin{equation}
    \widetilde{c}_j(k) \vert {\rm g.s.} \rangle = 0 \quad (j = 1, \ldots, n) .
\end{equation}
($n$ is the total number of bands.) Put differently, such ground states are in one-to-one correspondence with collections of complex vector spaces $\{ A_k \}_{k \in M}$ where we take
\begin{equation}
    A_k = \mathrm{span}_\mathbb{C} \{ \widetilde{c}_1(k), \ldots, \widetilde{c}_n(k) \}
\end{equation}
to be spanned by the quasi-particle annihilation operators at $k$; more precisely, by those \emph{lowering} the momentum by $k$.

To give an example, consider a system with conserved particle number (or charge) and two bands, one conduction and one valence band, labeled by $p$ and $h$ respectively. In this case the annihilation vector space $A_k$ is spanned by two operators: the one removing a particle in the conduction band at momentum $k$ and another one creating a particle in the valence band at $-k$:
\begin{equation}
    A_k = \mathrm{span}_\mathbb{C} \{ c_{k,\,p}\, , c_{-k,\,h}^\dagger \} .
\end{equation}
Here we have $n = 2$, $\widetilde{c}_1(k) \equiv c_{k,\,p}\,$, and $\widetilde{c}_2(k) \equiv c_{-k,\,h}^\dagger \,$. In the more general case of a superconductor, the quasi-particle annihilation operators are obtained by a Bogoliubov transformation
\begin{equation}
    \widetilde{c}_j (k) =
    \sum_{l=1}^n \left( u_{lj}(k)\, c_{k,\,l} + v_{lj}(k)\, c_{-k,\,l}^\dagger \right) ,
\end{equation}
with complex coefficients $u_{lj}(k)$ and $v_{lj}(k)$.

If the physical system were a metal, the vector space $A_k$ of quasi-particle annihi\-lation operators as a function of $k$ would jump at the Fermi surface, but since we are considering gapped systems, the assignment $k \mapsto A_k$ has the good feature of being continuous everywhere, giving a vector bundle $\{ A_k \}_{k \in M}$.

Now we observe that even in the absence of any symmetries (other than trans\-lations), the fibers $A_k$ are constrained by the condition that any pair of annihilation operators must have vanishing anti-commutator by the canonical anti-commutation relations for fermions. We write this condition summarily as
\begin{equation}
    \{ A_k \,, A_{-k} \} = 0 ,
\end{equation}
and refer to it as the Fermi constraint. Returning to the Periodic Table, let us point out that what is commonly known as the ``particle-hole symmetry'' of class $D$ is nothing but the Fermi constraint relating opposite fibers ($A_k$ with $A_{-k}$) in our model. Another comment directed at the experts is that our vector bundle $\{ A_k \}_{k \in M}$ is complex and cannot be viewed as a real vector bundle in any traditional sense. (Indeed, while the Fermi constraint determines $A_{-k}$ from $A_k$, it does not give rise to any complex-linear or anti-linear mapping between the individual vectors of these two spaces.) See, however, the notion of twisted vector bundle in \cite{FreedMoore}.

\subsection{Universal model (including symmetries)}\label{sect:3.2}

Our next step is to refine the model of translation-invariant free-fermion ground states as vector bundles by incorporating symmetry operations that commute with translations. (Note that we do not consider space group symmetries, thus excluding topological crystalline insulators). In that process, as outlined by Kitaev, true physical symmetries are converted into ``pseudo-symmetries''. Let us write down the outcome first and do the explaining afterwards.

For a system with symmetry index $s$, one is given a (representation of a) Clifford algebra with $s$ generators:
\begin{equation}\label{eq:Cliff}
    J_l J_m + J_m J_l = - 2 \delta_{l m} \mathbf{1} \quad (1 \leq l,m \leq s) ,
\end{equation}
where $J_1, \ldots, J_s$ are unitary operators on the sum of $A_k$ with its orthogonal complement, $A_k^\mathrm{c}\,$. Since that sum comprises all single-fermion operators (annihilators as well as creators), it is independent of the momentum: $A_k \oplus A_k^\mathrm{c} \equiv \mathbb{C}^{2n}$. All operators $J_1, \ldots, J_s$ preserve the canonical anti-commutation relations expressed by $\{ \, , \, \}$.

A concise summary of the outcome of incorporating symmetries is the following.

\smallskip\noindent\textbf{Definition. --} \textit{By the translation-invariant ground state of a gapped free-fermion system of symmetry class $s$ we mean a sub-vector bundle $\{ A_k \}_{k \in M}$ with fibers $A_k \subset \mathbb{C}^{2n}$ of rank $n = \mathrm{dim}\, A_k$ subject to (for all $k \in M$)}
\begin{eqnarray*}
    &\textit{1. Fermi constraint}: \; \{ A_k \,, \, A_{-k} \} = 0\,; \cr
    &\textit{2. Pseudo-symmetries}: \; J_1 A_k = \ldots = J_s \, A_k = A_k^\mathrm{c} \,.
\end{eqnarray*}
We speak of $J_1, \ldots, J_s$ as ``pseudo-symmetries'' because each of them sends $A_k$ to its orthogonal complement $A_k^\mathrm{c}\,$, whereas a true unitary symmetry would map $A_k$ to itself.

\smallskip\noindent\textbf{Example. --} Let the gapped system (and hence its ground state) be time-reversal invariant. Then we have $T A_k = A_{-k}$ or verbally: applying the anti-unitary time-reversal operator $T$ to any quasi-particle annihilation operator at momentum $k$ we get a quasi-particle annihilation operator at the opposite momentum $-k$. If we take the further step of applying the operator $\gamma$ of Hermitian conjugation ($\gamma : \; c \leftrightarrow c^\dagger$), we end up with a creation operator back at $+k$. This means that the composition $J_1 = \gamma T$ sends $A_k$ to its orthogonal complement $A_k^\mathrm{c}$, which yields the first pseudo-symmetry condition: $J_1 A_k = A_k^\mathrm{c}$. Moreover, being the product of two anti-unitary operators, $J_1$ is unitary; and assuming the case of fermions with half-integer spin, $T$ squares to minus one, $J_1$ does the same, and we have identified the first generator $J_1$ of the Clifford algebra of pseudo-symmetries.

\smallskip\noindent\textbf{Remark. --} As sub-vector bundles for a fixed ambient vector bundle (with fiber $\mathbb{C}^{2n}$), our vector bundles $\{ A_k \}_{k \in M}$ come with a natural notion of homotopy amongst them. The equivalence relation given by homotopy divides them into homotopy classes. It is the sets of these homotopy classes that are listed in Table \ref{fig:1}. Please be advised that the said homotopy classes do not form homotopy groups, i.e., there exists no natural notion of adding much less subtracting them (unless one approximates them by a $K$-theory construction). Thus the entries of Table \ref{fig:1} are fundamentally just sets, not groups.

\subsection{Kitaev sequence}

We have seen how the true anti-unitary symmetry $T A_k = A_{-k}$ gets transcribed to one pseudo-symmetry $J_1 A_k = A_k^\mathrm{c}$. How does this story continue? While Kitaev \cite{kitaev} wrote down only the first two steps, the following diagram presents the whole answer.
\begin{center}
\begin{tabular}{l|l|l|l}
class   &true symmetries &$s$    &pseudo-symmetries\\
\hline
$D$     &none		&$0$		&Fermi constraint\\
$\DIII$     &$T$ (time reversal)		&$1$		&$J_1 = \gamma \, T$\\
$\AII$     &$T, Q$ (charge)		&$2$		&$J_2 = \mathrm{i} \gamma \, T Q$ \\
$\CII$     &$T, Q$, $C$ (ph-conj.)		&$3$		&$J_3 = \mathrm{i} \gamma \, C Q$\\
\hline
$C$     &$S_1$, $S_2$, $S_3$ (spin rot.)		&$4$		&see text\\
$\CI$     &$S_1$, $S_2$, $S_3$, $T$		&$5$		&\\
$\AI$     &$S_1$, $S_2$, $S_3$, $T$, $Q$ 		&$6$		&\\
$\BDI$     &$S_1$, $S_2$, $S_3$, $T$, $Q$, $C$	&$7$		&\\
\end{tabular}
\end{center}
Remember that even if there are no symmetries (beyond translations), we still have the Fermi constraint due to Fermi statistics; this case, $s = 0$, is known as class $D$. As we have seen, imposing $T$ (with $T^2 = - \mathbf{1}$) gives one pseudo-symmetry with $J_1$ as generator, which puts us in the situation of $s = 1$, also known as the superconducting class $\DIII$. Next, we add the requirement that the particle number or charge (with operator $Q$ and $Q^2 = + \mathbf{1}$) be conserved. This allows us to form a second generator, $J_2 = \mathrm{i} Q J_1 = - \mathrm{i} J_1 Q$, which is readily seen to anti-commute with $J_1$ and square to minus the identity. In the present context, adding the true (unitary) symmetry of charge conservation, $Q A_k = A_k$, is equivalent to adding a second pseudo-symmetry, $J_2 A_k = A_k^\mathrm{c}$. We have now arrived at $s = 2$, a.k.a.\ class $\AII$. Next, to move on to $s = 3$, or class $\CII$, we include twisted particle-hole conjugation $C$ (with $C^2 = + \mathbf{1}$) as a third symmetry -- more precisely, as a true anti-unitary symmetry which commutes with the Hamiltonian. The additional physical symmetry $C A_k = A_{-k}$ translates into a third pseudo-symmetry condition, $J_3 A_k = A_k^\mathrm{c}$. To continue even further, we wipe the plate clean by erasing all symmetries $T$, $Q$, and $C$, and we demand instead that the spin-rotation generators $S_1$, $S_2$, $S_3$ be symmetries. The rest of the story is a repetition of what happened at the beginning.

It remains to explain why the 3 generators $S_1, S_2, S_3$ of the spin-rotation group together with $s-4$ symmetries (taken from $T, Q, C$) amount to $s$ pseudo-symmetries. To that end, we must invoke the so-called $(1,1)$-periodicity theorem, as follows. Let
\begin{equation}\label{eq:3.7}
    C_s(n) := \{ A \subset \mathbb{C}^{2n} \mid J_1 A = \ldots = J_s A = A^\mathrm{c} \}
\end{equation}
denote the so-called classifying space for class $s$, i.e.\ the space of all vector spaces $A \equiv A_k$ allowed by $s$ pseudo-symmetries (we suppress the index $k$ for now). Inside it, we have the subspace $R_s(n)$ of annihilation $n$-planes $A$ that also satisfy the Fermi constraint:
\begin{equation}\label{eq:3.8}
    R_s(n) := \{ A \in C_s(n) \mid \{ A , A \} = 0 \} .
\end{equation}
Let us mention in passing the celebrated 8-fold periodicity
\begin{equation}
    R_s(n) \simeq R_{s+8}(16n) ,
\end{equation}
which is (just) one of the mathematical phenomena behind the Periodic Table.

Now we double the number of bands, going from $n$ to $2n$ (or more formally, replacing $\mathbb{C}^{2n}$ by $\mathbb{C}^{2n} \oplus \mathbb{C}^{2n}$), and on the doubled space we introduce the operators
\begin{eqnarray}\label{eq:doubling}
    I = \left( \begin{array}{ll} 0 &\mathbf{1}_{2n} \\ -\mathbf{1}_{2n} &0 \end{array} \right) , \quad K = \mathrm{i} \left( \begin{array}{ll} \mathbf{1}_{2n} &0 \\ 0 &-\mathbf{1}_{2n} \end{array} \right) , \cr \widetilde{J}_l = \left( \begin{array}{ll} 0 &J_l \cr J_l &0 \end{array} \right) \quad (l = 1, \ldots, s).
\end{eqnarray}
This extends the Clifford algebra of $J_1, \ldots, J_s$ by two extra generators, $I$ and $K$, with one important subtlety: the distinguished generator $K$ reverses the sign of the anti-commutator bracket: $\{ K w , K w^\prime \} = - \{ w , w^\prime \}$, whereas all others preserve it. We call such a generator $K$ ``imaginary'', while $I$ and the $J$'s are called ``real''.

In this setting one has a bijection, or rather a pair of bijections, each of which relates the classifying space determined by the original Clifford algebra to the corresponding classifying space for the extended algebra:
\begin{equation}
    C_s(n) \simeq C_{s+2}(2n) , \quad R_s(n) \simeq R_{s+1,1}(2n) .
\end{equation}
In both instances, the bijection goes by
\begin{equation}\label{eq:bij-4-A}
    A \mapsto \widetilde{A} := \{ \Big( \begin{array}{l} w+w^\prime \cr w-w^\prime \end{array} \Big) \mid w \in A , \; w^\prime \in A^\mathrm{c} \} .
\end{equation}
(The proof is an exercise in linear algebra.) Thus if we double the number of bands and add two pseudo-symmetries, one real and one imaginary, then the situation remains unchanged. This is the content of the $(1,1)$-periodicity theorem.

It should be stressed that the statement of $(1,1)$ periodicity does not call for the generators to be realized in the explicit form of (\ref{eq:doubling}). Rather, all that matters are the Clifford algebra relations (\ref{eq:Cliff}) for $J_1, \ldots, J_s, I, K$ and the presence of one imaginary generator $K$. (The precise details are spelled out in \cite{RK-MZ}.) We will use this independence of the choice of basis for $\mathbb{C}^{2n} \oplus \mathbb{C}^{2n}$ when working through some examples later.

We are now in a position to settle the issue in question. Given the spin-rotation generators $S_1, S_2, S_3$ and $s-4$ pseudo-symmetries $J_5, \ldots, J_s$ we put the former on the diagonal blocks of the doubled space and the latter on the off-diagonal blocks:
\begin{equation}
    \widetilde{J}_l := \left( \begin{array}{ll} \mathrm{i} S_l &0 \\ 0 &-\mathrm{i} \mathrm{S}_l \end{array} \right) \quad (l \leq 3), \quad \widetilde{J}_l := \left( \begin{array}{ll} 0 &J_l \\ J_l &0 \end{array} \right) \quad (l \geq 5) .
\end{equation}
We also identify $\widetilde{J}_4 \equiv I$ and note that $K = \mathrm{i} \widetilde{J}_1 \widetilde{J}_2 \widetilde{J}_3$.

Now, on physical grounds, the spin-rotation generators commute with the pseudo-symmetries ($J_5, J_6, J_7$) drawn from ($T, Q, C$). It follows directly that the new operators $\widetilde{J}_1, \ldots, \widetilde{J}_s$ satisfy the standard Clifford algebra relations, and by an easy argument using $(1,1)$ periodicity one verifies that the $s$ pseudo-symmetries $\widetilde{J}_1, \ldots, \widetilde{J}_s$ are indeed equivalent to $s-4$ pseudo-symmetries $J_5, \ldots, J_s$ in conjunction with the three spin-rotation symmetries $S_1$, $S_2$, and $S_3$. A crucial point here is that, once again, the true symmetries $S_l A = A$ get transformed into pseudo-symmetries $\widetilde{J}_l \widetilde{A} = \widetilde{A}^\mathrm{c}$ ($l = 1, 2, 3$).

In summary, there exists a first-principles reason why one wants to put the eight classes of the ``real'' sub-table in the order of $s = 0, 1, 2, 3, 4, 5, 6, 7$ or equivalently, $D, \DIII, \AII, \CII, C, \CI, \AI, \BDI$.

\subsection{Disorder}

Having assumed translation invariance so far, let us now say a few words about disordered systems. Disorder of course kills the law of momentum conservation, and one might worry that all the formalism based on conserved momentum goes down the drain. However, this is not so. From the work of Bellissard and others on the integer quantum Hall effect, one knows in fact what to do. Adopting the non-commutative geometry setting of Connes, one turns to the C$^\ast$-algebra of bounded observables and exploits the integral pairing of its $K$-groups with cyclic cohomology as given by the Chern--Connes character; see \cite{prodan} for a recent review. This approach has already been adapted \cite{p-hsb} to some parts of the Periodic Table.

If one wishes to stay as close as possible to the framework outlined here, one may boost the system by a momentum translation through $k$ and consider the vector bundle $\{ A_k \}_{k \in M}$ of disordered ground states (for any fixed disorder realization) over the manifold $M$ of boost parameters $k$. In this modified setting our algebraic framework of Fermi constraint and pseudo-symmetries (from true symmetries) remains intact. This remark is crucial for the grand perspective, as it is the stability with respect to disorder that singles out the $8 + 2 = 10$ classes of the Periodic Table from the plethora of topological crystalline insulators.

This brings us to the question: why are there exactly ten classes in the disordered setting and not more? The answer is well documented (see \cite{symclasses} and the references given therein), so let us just say this. It is a theorem, dubbed the ``Tenfold Way'', that no matter what group of unitary and anti-unitary transformations you start from -- be it discrete or continuous -- you must always end up with a classifying space of one of the ten known types, provided that your symmetries are true physical ones. There is no conflict with  the existence of topological crystalline insulators, as the latter arise from the possibility of space group symmetries causing relations among the fibers $A_k$ for different momenta $k$, while the underlying classifying space remains the same.

\subsection{Complex classes}

Of course, to reach the count of ten one augments the eight ``real'' classes discussed above by two ``complex'' ones:
\begin{center}
\begin{tabular}{l|l|l}
class   &symmetries &pseudo-syms\\
\hline
$A$
& $Q$
& none\\
$\AIII$
&$Q$,
$C$	
&$J_1 = \mathrm{i} \gamma \, C$
\end{tabular}
\end{center}
To realize class $A$, one takes the charge ($Q$) to be conserved. This effectively cancels the Fermi constraint and one is left with no pseudo-symmetries at all. Ground states in this case are just plain complex vector bundles. Finally, by imposing an additional symmetry of particle-hole conjugation, one arrives at class $\AIII$, featuring one pseudo-symmetry.

\section{The Diagonal Map}

So much for symmetries and the universal model. Before we get to our main point, the Diagonal Map, we wish to make another remark about the basic principles of the game.

\subsection{How to classify?}

There exist several notions of topological equivalence for vector bundles, and they are not the same. The finest classification is by homotopy -- two of our vector bundles belong to the same homotopy class if they are adiabatically connected, i.e., can be transformed into each other by a continuous sequence of infinitesimal deformations. Another classification is by isomorphy -- two vector bundles are said to be isomorphic if an isomorphism (not necessarily connected to the identity) takes one into the other. Isomorphy is coarser than homotopy in general. Indeed, two vector bundles in the same isomorphism class need not be in the same homotopy class, unless the number of conduction bands is sufficiently large. Third, if not only the conduction bands but also the valence bands are taken to be very numerous, isomorphism classes stabilize as so-called $K$-theory classes. (It should in principle be discussed which of these mathematical notions is most appropriate to the physics of topological insulators. Little discussion of this issue seems to exist in the current literature.)

To give an example, three-dimensional spin-singlet superconductors with broken time-reversal symmetry ($s = 4$, or class $C$) support one topologically non-trivial phase, provided that the system is described by a single band ($n = 2$). This is predicted by the homotopy classification of \cite{RK-MZ}, even though the corresponding entry of the Periodic Table vanishes. Such discrepancies are to be expected, as homotopy groups are known to be vastly more intricate than $K$-theory groups. To give another example, for systems of class $D$ with two momentum-like and one space-like dimension (the latter associated with a defect), the $K$-theory treatment of Teo and Kane predicts a $\mathbb{Z}_2$-classification. While this is the correct answer in the $K$-theory limit of many bands, it is not correct for small $n$. In fact, the minimal case $(n = 1)$ can be shown to be classified by $\mathbb{Z}$.

\subsection{Diagonal Map $(d,s) \to (d+1,s+1)$}

We now sketch the Diagonal Map, taking a $d$-dimensional ground state of class $s$ and turning it into a $(d+1)$-dimensional ground state of class $s+1$. For this purpose, we assume the momentum space to be a sphere: $M = \mathrm{S}^d$. While this assumption is not necessary (see \cite{RK-MZ} and \cite{KG14} for the general case), it does simplify the discussion.

So let there be some ground state (or vector bundle) of class $s$. It turns out that the workings of the Diagonal Map require one imaginary pseudo-symmetry. Since the standard setting of class $s$ as laid down in the Definition of Section \ref{sect:3.2} does not provide for such an object, we help ourselves by using $(1,1)$ periodicity. Thus we jack up the given data by doubling the number of bands, extending the Clifford algebra of pseudo-symmetries by two generators, $I$ and $K$, one of which is imaginary ($K$), and lifting the vector bundle isomorphically to the doubled band space. After this, we have $s+2$ pseudo-symmetries $J_1, \ldots, J_s, I, K$ and a vector bundle $\{ A_k \}_{k \in M}$ with rank-$2n$ fibers in $\mathbb{C}^{2n} \oplus \mathbb{C}^{2n}$ (to simplify the notation, we have dropped the tilde of before).

Given these initial data, we manufacture a derived vector bundle of class $s + 1$ over $M = \mathrm{S}^{d+1}$ as follows. Let $t \in [-\pi/2,\pi/2]$ be a coordinate for the extra dimension to be added. (More precisely, we think of $t$ as a polar coordinate for the embedding of $\mathrm{S}^d$ as an equator into $\mathrm{S}^{d+1}$.) Then we give a $t$-dependence to the fibers $A_k$ by applying a one-parameter group of unitary transformations:
\begin{equation}\label{eq:4.1}
    A_{k,t} := \mathrm{e}^{(t/2) K J(A_k)} \cdot A_k ,
\end{equation}
where $J(A) = \mathrm{i} (\Pi_A - \Pi_{A^\mathrm{c}})$ is the operator that multiplies by $\mathrm{i}$ on $A$ and by $-\mathrm{i}$ on the orthogonal complement $A^\mathrm{c}$. By investing the algebraic properties at hand, one verifies that this definition has all the right properties:
\begin{enumerate}
\item
the Fermi constraint is satisfied:
\begin{equation}
    \{ A_{k,\,t}\, , \, A_{-k,-t} \} = 0 \,;
\end{equation}
\item
the pseudo-symmetry conditions hold:
\begin{equation}
    J_1 A_{k,t} = \ldots = J_s A_{k,t} = A_{k,t}^\mathrm{c} = I A_{k,t} \,;
\end{equation}
\item
and the fibers become $k$-independent at $t = \pm \pi/2$:
\begin{equation}
    A_{k, t = \pm \pi/2} = E_{\mp \mathrm{i}}(K) ,
\end{equation}
where $E_\lambda(K)$ denotes the eigenspace of $K$ with eigenvalue $\lambda$.
\end{enumerate}
By the third property, the assignment $(k,t) \mapsto A_{k,t}$ for $t \in [0,\pi/2]$ can be viewed as a function on the northern hemisphere $\mathrm{S}^d \times [0,\pi/2] \equiv D_+^{d+1}$, and the same goes for $t \leq 0$ (southern hemisphere $D_-^{d+1}$). Thus we have extended the initial vector bundle $\{ A_k \}$ over the momentum sphere $\mathrm{S}^d$ to a final vector bundle $\{ A_{k,t} \}$ over the momentum sphere $\mathrm{S}^{d+1} = D_+^{d+1} \cup D_-^{d+1}$ with polar coordinate $t \in [-\pi/2,\pi/2]$. In the process we added one pseudo-symmetry given by the real generator $I$. The finished product is a vector bundle $\{ A_{k,t} \}$ of class $s+1$ in dimension $d+1$.

Let us illustrate this procedure by a few informative examples.

\subsection{Example 1}

As a first example, we start from non-trivial data for class $\BDI$ in dimension zero and apply the Diagonal Map to manufacture a non-trivial vector bundle of class $D$ in one dimension (the so-called Majorana chain). Introduced as the class with seven real pseudo-symmetries, $\BDI$ is realized more easily by means of a single imaginary pseudo-symmetry. We will utilize this alternative realization, which is offered by the equivalence $R_7(8n) \simeq R_{8,1}(16n) \simeq R_{0,1}(n)$ due to $(1,1)$ periodicity in combination with 8-fold periodicity. The imaginary pseudo-symmetry of class $\BDI$ is $K_1 = \mathrm{i} \gamma T$ for a time-reversal operator $T$ with $T^2 = + \mathbf{1}$.

Assuming $n = 1$, we have a classifying space $R_{0,1}(1)$ consisting of just two points, the complex line of $c$ (empty state) and that of $c^\dagger$ (occupied state):
\begin{equation}
    R_{0,1}(1) =  \{ \mathbb{C} \cdot c \, , \, \mathbb{C} \cdot c^\dagger \} .
\end{equation}
More generally, $R_{0,1}(n) \simeq \mathrm{O}_n$ consists of two connected components distinguished by even versus odd fermion parity (or the sign of the determinant on $\mathrm{O}_n$).

For simplicity, we restrict the discussion to the case of $n = 1$. We assign the line $A \equiv \mathbb{C} \cdot c^\dagger$ to the eastern pole (or $k = 0$) of $M = \mathrm{S}^0 \simeq \{ 0, \pi \}$ and the opposite line $\mathbb{C} \cdot c$ to the western pole (or $k = \pi$). In applying the Diagonal Map in the present case, we have license to simply skip the doubling process $n \to 2n$ of $(1,1)$ periodicity, as there is already one imaginary pseudo-symmetry $K_1$ at our disposal. By using the relations
\begin{equation}
    J(A) c^\dagger = \mathrm{i} c^\dagger , \quad J(A) c = - \mathrm{i} c , \quad K_1 c^\dagger = \mathrm{i} c , \quad K_1 c = \mathrm{i} c^\dagger ,
\end{equation}
and letting $t \equiv k$ run through the interval $[-\pi/2, \pi/2]$, we compute
\begin{equation}\label{eq:4.7}
    A_k = \mathrm{e}^{(k/2) K_1 J(A)} \cdot A = \mathbb{C} \cdot \big( c_{-k}^\dagger \cos(k/2) - c_k \sin(k/2) \big) ,
\end{equation}
which defines the fibers $A_k$ over half of the circle $\mathrm{S}^1$. To compute those over the other half, we replace the line $\mathbb{C} \cdot c^\dagger$ by the line $\mathbb{C} \cdot c$. Joining the two semicircles together, we still get the formula (\ref{eq:4.7}), but now with $k$ running over the full Brillouin zone $[-\pi,\pi]$.

The vector bundle (\ref{eq:4.7}) corresponds to a special representative of the Majorana chain. Written in BCS form, the ground state annihilated by $\{ A_k \}$ reads
\begin{equation}
    \vert {\rm BCS} \rangle = \mathrm{e}^{\frac{1}{2} \sum_k \cot(k/2) \, c_k^\dagger c_{-k}^\dagger} \vert \mathrm{vac} \rangle .
\end{equation}
It is not homotopic to the trivial vacuum state because it has odd (even) fermion parity at $k = 0$ (resp.\ $k = \pi$) and this property is a topological invariant even in the absence of any symmetries ($s = 0$).

\subsection{Example 2}

For a second example, we start from the outcome of the previous one ($s = 0$, $d = 1$) and progress to a two-dimensional superconductor with time-reversal invariance ($s = 1$, or class $\DIII$). The doubling procedure of $(1,1)$ periodicity here amounts to forming the tensor product with two-dimensional spin space, $(\mathbb{C}^2)_\mathrm{spin}$. Thus we now have four types of single-fermion operator: $c_\uparrow, c_\downarrow, c_\uparrow^\dagger, c_\downarrow^\dagger$. The real generator $I$ is to be identified with the first pseudo-symmetry of the Kitaev sequence:
\begin{equation}
    \hspace{-1cm}
    I \equiv J_1 = \gamma T , \quad I c_\downarrow = c_\uparrow^\dagger \,, \quad I c_\uparrow^\dagger = - c_\downarrow \,, \quad I c_\uparrow = - c_\downarrow^\dagger \,, \quad I c_\downarrow^\dagger = c_\uparrow \,.
\end{equation}
(Recall $T^2 = - \mathbf{1}$ for class $\DIII$.) A good choice of imaginary generator $K$ is
\begin{equation}
    \hspace{-1cm}
    K c_\downarrow = \mathrm{i} c_\uparrow^\dagger \,, \quad K c_\uparrow^\dagger = \mathrm{i} c_\downarrow \,, \quad K c_\uparrow = \mathrm{i} c_\downarrow^\dagger \,, \quad K c_\downarrow^\dagger = \mathrm{i} c_\uparrow \,.
\end{equation}
Note that $IK + KI = 0$ and $K^2 = - \mathbf{1}$. For this choice of $K$, the vector bundle (\ref{eq:4.7}) of the Majorana chain $(1,1)$-doubles to
\begin{equation}
    \hspace{-1cm}
    A_k = \mathrm{span}_\mathbb{C} \{ \widetilde{c}_+(k) , \widetilde{c}_-(k) \}
\end{equation}
with
\begin{equation}
    \widetilde{c}_\pm (k) = c_{-k, \pm}^\dagger \cos(k/2) - c_{k,\mp} \sin(k/2) , \quad c_{\pm} = (c_\uparrow \pm c_\downarrow)/\sqrt{2} .
\end{equation}
Indeed, one easily checks that the pseudo-symmetry conditions $K A_k = I A_k = A_k^\mathrm{c}$ are obeyed. We then apply the one-parameter group of the Diagonal Map to obtain
\begin{equation}\label{eq:4.13}
    A_{\bf k} = \mathrm{e}^{(k_2/2) K J(A_{k_1})} \cdot A_{k_1} =
    \mathrm{span}_\mathbb{C} \left\{ \widetilde{c}_+({\bf k}) ,
    \widetilde{c}_-({\bf k}) \right\} ,
\end{equation}
where ${\bf k} = (k_1 , k_2)$ and
\begin{eqnarray}\label{eq:4.14}
    \widetilde{c}_\pm ({\bf k}) &&= \big( c_{-{\bf k},\pm}^\dagger \cos(k_1/2) - c_{{\bf k},\mp} \sin(k_1/2) \big) \cos(k_2/2) \cr
    &&+ \big( \mp c_{-{\bf k},\mp}^\dagger \sin(k_1/2) \mp c_{{\bf k},\pm} \cos(k_1/2) \big) \sin(k_2/2) .
\end{eqnarray}
By writing the ground state in BCS form, one sees that this is a superconductor with spin-triplet pairing. (The Cooper-pair state is anti-symmetric in $k$-space and symmetric in spin space.) We assert that it is in a symmetry-protected topological phase, as the winding in its ground state cannot be undone without breaking the time-reversal invariance.
Let us verify this assertion.

By using the pseudo-symmetry $J_1$, one defines on $\mathrm{span}_\mathbb{C} \{ c_\uparrow, c_\downarrow, c_\uparrow^\dagger, c_\downarrow^\dagger \} \equiv \mathbb{C}^4$ (we have set $n = 2$ here, but the same reasoning would go through for $\mathbb{C}^{2n}$ with general $n$) a complex bilinear form $\omega$ as
\begin{equation}
    \omega(v,v^\prime) = \{ J_1 v , v^\prime \} .
\end{equation}
Since the unitary generator $J_1$ satisfies $J_1^2 = - \mathbf{1}$ and is real (i.e., preserves $\{ \, , \, \}$), this bilinear form is skew-symmetric:
\begin{equation}
    \omega(v,v^\prime) = \{ J_1^2 v , J_1 v^\prime \} = - \{ v , J_1 v^\prime \} = - \omega(v^\prime,v) .
\end{equation}
Moreover, $\omega$ is non-degenerate on $\mathbb{C}^4$, because so is $\{ \, , \, \}$ and $J_1$ has an inverse. Note that by the Fermi constraint $\{ A_{-{\bf k}} \,, A_{\bf k} \} = 0$ and the pseudo-symmetry condition $J_1 A_{\bf k} = A_{\bf k}^\mathrm{c}$ the bilinear form $\omega$ remains non-degenerate when restricted to $A_{-{\bf k}} \otimes A_{\bf k}\,$. Thus $A_{\bf k}$ is paired via $\omega$ with $A_{-{\bf k}}\,$; we also say that $A_{-{\bf k}}$ is the $\omega$-dual of $A_{\bf k}\,$.

The relevant object now is the skew-symmetric form $\omega$ restricted to the fibers $A_{\bf k}$:
\begin{equation}
    \omega_{\bf k} := \omega \big\vert_{A_{\bf k} \otimes A_{\bf k}}\,.
\end{equation}
There exists no guarantee for this restriction to be non-degenerate. (In order to have a non-trivial restriction, one drops the ${\bf k}$-dependence of the single-fermion operators: $c_{{\bf k},\,\sigma} \to c_\sigma \,$, etc.) In fact, the homotopy class of the vector bundle $\{ A_{\bf k} \}$ is diagnosed by counting the zeroes of the Pfaffian of $\omega_{\bf k}\,$. More precisely, it is the even-odd parity of the number of pairs of zeroes of $\mathrm{Pf}(\omega_{\bf k})$ that is invariant. This fact derives from the following properties (c.f.\ \cite{KaneMele} for the $\AII$ case). (i) By the relations $\overline{\omega_{\bf k}(v,v^\prime)} = \omega_{-{\bf k}}(Tv , Tv^\prime)$ and $T A_{\bf k} = A_{-{\bf k}}$ the zeroes occur in pairs $({\bf k} , - {\bf k})$. (ii) Because $A_{-{\bf k}}$ is the $\omega$-dual of $A_{\bf k}\,$, no zeroes can occur at $T$-invariant momenta ${\bf k} = - {\bf k}\,$. (iii) As zeroes of a complex-valued function in two dimensions, the zeroes of $\mathrm{Pf}(\omega_{\bf k})$ carry a vorticity. (iv) Under continuous deformations of the vector bundle, only quadruples of zeroes can be created or annihilated (by the creation or annihilation of vortex-antivortex pairs).

Our search for zeroes of $\mathrm{Pf} (\omega_{\bf k})$ is simplified by the observation that (for $n = 2$) such zeroes occur if and only if $A_{\bf k}$ is a Lagrangian subspace of $\mathbb{C}^4$:
\begin{equation}
    \omega_{\bf k}(v,v^\prime) = 0 \quad {\rm for \; all} \; v, v^\prime \in A_{\bf k}\,.
\end{equation}
This condition can be reformulated as $A_{\bf k}^\mathrm{c} = A_{-{\bf k}}$ (``band inversion''), i.e.\ the quasi-particle creation operators at momentum ${\bf k}$ must be the quasi-particle annihilators at $-{\bf k}\,$. Inspection of (\ref{eq:4.14}) reveals that $A_{{\bf k}_\ast}^\mathrm{c} = A_{- {\bf k}_\ast}$ for ${\bf k}_\ast = (k_0 , k_1) = (\pi/2, \cdot)$. Indeed,
\begin{equation}
    A_{{\bf k}_\ast} = \mathrm{span}_\mathbb{C} \left\{c_{-{\bf k}_\ast , +}^\dagger - c_{\,{\bf k}_\ast , +}^{\vphantom{\dagger}}\,, c_{-{\bf k}_\ast , -}^\dagger + c_{\,{\bf k}_\ast , -}^{\vphantom{\dagger}} \right\} = A_{-{\bf k}_\ast}^\mathrm{c} .
\end{equation}
The points $\pm {\bf k}_\ast$ are the only points where this happens. Thus the ground-state vector bundle (\ref{eq:4.13}) is in a topological phase with non-trivial $\mathbb{Z}_2$ (or Kane-Mele) invariant. As we have seen, this topological phase originates (via the Diagonal Map) from the Majorana chain as its direct ancestor.

\subsection{Example 3}

For a third example, we might start from the outcome of the previous one and progress to a three-dimensional band insulator in class $\AII$. Here the effect of doubling by $(1,1)$ periodicity would be to introduce two bands (one conduction and one valence) for the already spinful system. The very same step ($\DIII \to \AII$) but from $d=1$ to $d=2$ was spelled out in detail in \cite{RK-MZ}. Let us therefore turn to another example for more variety.

We start from data for class $\AI$ in dimension zero and pass to a superconductor of class $\BDI$ in dimension one (a.k.a.\ the Kitaev chain). By an argument that was already used in Example 1 above, we may realize class $\AI$ (with $s = 6$ real generators) by means of two imaginary generators $K_1$ and $K_2$ (or $s = -2$). These are expressed by
\begin{equation}
    K_1 = \mathrm{i} \gamma T = \mathrm{i} T \gamma, \quad K_2 = \mathrm{i} Q K_1 = - \mathrm{i} K_1 Q ,
\end{equation}
where $Q$ is the charge operator of before, and $T$ is time reversal for spinless particles ($T^2 = + \mathbf{1}$). Note that the standard Clifford algebra relations ($K_1^2 = K_2^2 = - \mathbf{1}$ and $K_1 K_2 + K_2 K_1 = 0$) are satisfied. The pseudo-symmetry conditions for data of class $\AI$ in dimension zero are then
\begin{equation}
    K_1 A = K_2 A = A^\mathrm{c} .
\end{equation}
In keeping with our general framework based on the Fermi constraint $\{ A , A \} = 0$, these are equivalent to the true symmetries of time reversal ($T A = A$) and charge conservation ($Q A = A$). The latter implies that any vector in the solution space $A$ must be either a single-fermion annihilation operator ($c$) or a single-fermion creation operator ($c^\dagger$), as it is these that span the two eigenspaces of $Q = - \mathrm{i} K_1 K_2$. Thus in a situation with $n$ bands we get to choose a decomposition $\mathbb{C}^n = V_- \oplus V_+$ into an empty/conduction subspace $V_-$ (annihilation by $c$) and an occupied/valence subspace $V_+$ (annihilation by $c^\dagger$). The space of such decompositions organizes into connected components labeled by $n_+ = \mathrm{dim} V_+$ for $0 \leq n_+ \leq n$. The number of such components is $n+1$ (which is the cardinality $2r+1 \equiv n+1$ that appears in row $s = 6$ and column $d = 0$ of Table \ref{fig:1}).

Now, to make a one-dimensional superconductor of class $\BDI$ we assign the trivial vacuum $(A_\ast = V_- = \mathbb{C}^n$ and $V_+ = 0$) to the point $k = \pi \in \mathrm{S}^0$ and some non-trivial data $A = V_- \oplus (V_+)^\ast$ to $k = 0 \in \mathrm{S}^0$. The latter are written as
\begin{equation}
    A = \mathrm{span}_\mathbb{C} \{ c_{-,i} \}_{i = 1, \ldots, n_-} \oplus
    \mathrm{span}_\mathbb{C} \{ c_{+,j}^\dagger \}_{j = 1, \ldots, n_+ } ,
\end{equation}
where $i = 1, \ldots, n_-$ ($j = 1, \ldots, n_+$) labels a basis for the conduction (resp.\ valence) subspace. By running the one-parameter group of the Diagonal Map $A_k = \mathrm{e}^{(k/2) K_2 J(A)} \cdot A$ for $k \in [-\pi/2,\pi/2]$ we then obtain
\begin{eqnarray}
    A_k &=& \mathrm{span}_\mathbb{C} \{ c_{-,i} \cos(k/2) + c_{-,i}^\dagger \sin(k/2) \}_{i = 1, \ldots, n_-} \cr &\oplus& \mathrm{span}_\mathbb{C} \{ c_{+,j}^\dagger \cos(k/2) + c_{+,j} \sin(k/2) \}_{j = 1, \ldots, n_+ } .
\end{eqnarray}
On the other hand, by running it with the vacuum data $A_\ast$ instead of $A$ and assigning the outcome to the complementary interval $\pi/2 \leq |k| \leq \pi$ about $k = \pi$, we get
\begin{equation}
    A_k = \mathrm{span}_\mathbb{C} \{ c_i \cos(k/2) + c_i^\dagger \sin(k/2) \}_{i = 1, \ldots, n} .
\end{equation}
Thus defined, the fibers $A_k$ depend continuously on $k$ at $k = \pm \pi/2$. By construction, they satisfy the Fermi constraint $\{ A_k , A_{-k} \} = 0$ and the pseudo-symmetry condition $K_1 A_k = A_k^\mathrm{c}$, as is required for a time-reversal invariant superconductor of class $\BDI$.

In the present case, the Diagonal Map cannot induce a bijection between homotopy classes. Indeed, we have seen that the $d = 0$ data of class $\AI$ fall into $n+1$ connected components, whereas systems of class $\BDI$ in one dimension are classified by the integers (c.f.\ Table \ref{fig:1}). The induced map here is an injection of $\mathbb{Z}_{n+1}$ into $\mathbb{Z}$. It becomes surjective in the $K$-theory limit of $n \to \infty$.

\section{Discussion}

The question now is whether the Diagonal Map between our free-fermion ground state vector bundles yields a bijection of homotopy classes or, physically speaking, a one-to-one correspondence of symmetry-protected topological phases along the diagonal of the Periodic Table for $d \geq 1$. The answer is yes, albeit with a trivial modification related to keeping base points fixed, and under two provisions: the $d$-dimensional momentum space $M$ must be a path-connected $\mathbb{Z}_2$-CW complex, adding one dimension is to be understood as leading to the $(d+1)$-dimensional suspension $\tilde{S} M$, and the number $n$ of bands cannot be too small in comparison with $d$. A rigorous proof is given in \cite{RK-MZ}.

Let us finish with a few words of perspective. The starting point of Bott's work \cite{Bott1959} was the insight that the space of minimal geodesics between antipodal points of a compact symmetric space $X$ is another compact symmetric space, $X^\prime$. Parametrizing geodesics by their midpoints, this gives a natural inclusion $X^\prime \hookrightarrow X$. By iterating this inclusion, Bott got two sequences of symmetric spaces, which in our notation read
\begin{equation}
    \ldots \hookrightarrow C_{s+1}(n) \hookrightarrow C_s(n) \hookrightarrow C_{s-1}(n) \hookrightarrow \ldots ,
\end{equation}
and
\begin{equation}
    \ldots \hookrightarrow R_{s+1}(n) \hookrightarrow R_s(n) \hookrightarrow R_{s-1}(n) \hookrightarrow \ldots .
\end{equation}
Moreover, by generalizing Morse theory to allow for degenerate critical points, he showed
that the space $X^\prime$ is a good approximation to the loop space of $X$ in the sense that their low-dimensional homotopy groups agree. In this way he computed the stable homotopy groups for all spaces $R_s(n)$ and $C_s(n)$. These turned out to be either $\mathbb{Z}$, or $\mathbb{Z}_2$, or $0$, which happen to be the entries occurring in the $d \geq 1$ part of the Periodic Table.

Yet, a second substantial insight was needed in order to arrive at Table \ref{fig:1}. It is not the homotopy groups of the symmetric spaces $C_s(n)$ and/or $R_s(n)$ that classify the symmetry-protected topological phases of gapped free fermions with disorder. Rather, the appropriate mathematical model is that based on the Kitaev sequence above. In our formulation, the additional structure (beyond Bott's setting) is the Fermi constraint due to the canonical anti-commutation relations. In the presence of $s$ pseudo-symmetries these yield a Fermi involution
\begin{equation}
    \tau_s : \; C_s(n) \to C_s(n) , \quad A \mapsto A^\perp ,
\end{equation}
with $A^\perp$ determined from $A$ by $\{ A^\perp , A \} = 0$. Its set of fixed points is $\mathrm{Fix}(\tau_s) = R_s(n)$. The Diagonal Map defined by (\ref{eq:4.1}) constructs from a point $A \in C_s(n) \simeq C_{s+2}(2n)$ a minimal geodesic in $C_{s+1}(2n)$. This construction is Fermi-equivariant, which is to say that it intertwines the action of $\tau_s$ on $C_s(n)$ with the action of $\tau_{s+1}$ on the loop space $\Omega \, C_{s+1}(2n)$ while reversing (a crucial difference from Bott!) the orientation of the loop. In view of Bott's scheme based on Morse theory, one expects the pair $(C_s(n), \tau_s)$ to be a good approximation to the pair $(\Omega\, C_{s+1}(2n) , \tau_{s+1})$ in the sense that the low-dimensional topology is captured. This expectation turns out to be true, and by recognizing the homotopy classes of vector bundles $\{ A_k \}_{k \in M}$ of symmetry class $s$ as the homotopy classes of $\tau$-equivariant classifying maps $M \to C_s(n)$, $k \mapsto A_k$, one arrives at Table \ref{fig:1}.

While we wish to advertise the homotopy-theoretic proof given in \cite{RK-MZ}, let it be stressed that all of the credit for identifying and assembling the relevant mathematical structures goes to Kitaev \cite{kitaev}. With this appreciation in mind, we suggest that Table \ref{fig:1} (including also the ``complex'' counterpart) be called the Bott-Kitaev Periodic Table.

\bigskip\noindent\textbf{Acknowledgment.} -- Financial support from the DFG via SFB/TR 12 is acknow\-ledged. The senior author is supported by DFG grant ZI 513/2-1, the junior author by a scholarship of the Deutsche Telekom Stiftung and a stipend of the Bonn-Cologne Graduate School of Physics \& Astronomy.

\section*{References}

\end{document}